# Dependencia de la temperatura de compensación en modelos de multicapas ferrimagnéticas de tipo 3($S_{1-x}$ $\sigma_x$)/$\sigma$

F.A. Dalmagro[1] y N. Hurtado[2]*

[1]*Laboratorio de Fenómenos No-Lineales, CEFITEC, Escuela de Física, UCV.*
[2]*Laboratorio de Física Teórica de Sólido, CEFITEC, Escuela de Física, UCV*

**Resumen**

En este trabajo hemos hecho un estudio numérico del comportamiento de la temperatura de compensación, $T_{comp}$, en multicapas ferrimagnéticas de tipo 3($S_{1-x}$ $\sigma_x$)/$\sigma$ con respecto al aumento de la concentración de espines $\sigma_x$, los cuales son distribuidos aleatoriamente en la mezcla. El modelo utilizado es el modelo de Ising con espines $S = 0,\pm 1$ y $\sigma = \pm 1/2$, con interacciones de intercambio ferromagnética entre espines del mismo tipo y antiferromagnética entre espines distintos. Hemos encontrado que la $T_{comp}$ disminuye lentamente con el aumento de $x$, hasta que $\sigma_x$ representa un poco más del 40% de los espines de la mezcla. Para valores superiores a este porcentaje, la $T_{comp}$ cae abruptamente y desaparece como lo predicen trabajos experimentales.

**Palabras clave:** Capas mixtas; ferrimagnetismo; ising; mezclas binarias; multicapas magnéticas; temperatura de compensación.

# Dependence of the compensation temperature in models of ferrimagnetic multilayers of the type 3($S_{1-x}$ $\sigma_x$)/$\sigma$

**Abstract**

We investigate numerically the behavior of the compensation temperature, $T_{comp}$, in ferrimagnetic multilayer in the type 3($S_{1-x}$ $\sigma_x$)/$\sigma$ with respect to the increase of the spin concentration $\sigma_x$, which are randomly distributive in a mix of spins $S = 0,\pm 1$ and $\sigma = \pm 1/2$, with ferromagnetic interaction between spins of the same type, and antiferromagnetic interaction for spins of different type. We have found that $T_{comp}$ decrease slowly with the increase of $x$, until $\sigma_x$ represent a little more than 40% of the spin mixture. For higher concentration values $T_{comp}$ drops abruptly and disappears, as predicted by experimental result.

0

## Introducción

Se ha encontrado experimentalmente que el acoplamiento entre los cristales *Gd/Co* es muy difícil de obtener (1) debido a la fuerte interdifusión entre las capas que toman lugar en el sistema, lo cual limita las posibles aplicaciones de estos sistemas ferrimagnéticos artificiales. Sin embargo, trabajos experimentales nos indican que con la preparación de mezclas amorfas de $Gd_{1-x}Co_x$, se produce una disminución tanto en la oxidación de la mezcla, así como en la velocidad de interdifusión en la interface (2). Con el modelo de interacción de intercambio no uniforme $Gd_{1-x}Co_x$, se ha logrado explicar algunas características del ordenamiento ferrimagnético de multicapas *Gd/Co* (3) En este modelo, el comportamiento macroscópico que se observa es el de un material ferrimagnético, en el cual la temperatura de compensación disminuye abruptamente con el espesor de la mezcla, como se observaba en el caso de las multicapas *Gd/Co*, sin embargo, el perfil magnético de $Gd_{1-x}Co_x$ es diferente al esperado para el caso ideal de *Gd/Co*. Numéricamente se han logrado reproducir algunas características del crecimiento de la capa $Gd_{1-x}Co_x$ (4). Estudios realizados con difracción de rayos-*X* en multicapas *Gd/Co* indican un rápido cambio al estado amorfo en la interface, especialmente cuando el *Co* es depositado sobre el *Gd* (5).

El objetivo principal de este trabajo es hacer un estudio numérico, del comportamiento de la temperatura de compensación en multicapas ferrimagnéticas del tipo $3(S_{1-x}\sigma_x)/\sigma$. En particular se usará el modelo de interacción de intercambio no uniforme (2) en este modelo. Se usa un modelo ferrimagnético mixto de Ising con una distribución aleatoria y uniforme de espines, para la capa mixta $S_{1-x}\sigma_x$, y el modelo de Ising ferromagnético para la capa pura $\sigma_x$. Se determina el comportamiento de la temperatura de compensación, $T_{comp}$, y la saturación de la magnetización al aumentar la concentración de espines $\sigma_x$.

## Modelo

Con la finalidad de reproducir algunos de los resultados observados experimentalmente en multicapas del tipo $Gd_{1-x}Co_x/Co$, hemos utilizado de forma combinada un modelo tridimensional ferrimagnético mixto de Ising para la capa mixta $Gd_{1-x}Co_x$, que llamaremos $S_{1-x}\sigma_x$, y el modelo ferromagnético mixto de Ising tridimensional para la capa $\sigma$. La muestra total consta de un 75% de capa mixta y un 25% de capa pura, $3(S_{1-x}\sigma_x)/\sigma$. En la ecuación [1] escribimos el Hamiltoniano que representa el modelo,

$$H = -J^{S-\sigma}_{<NN>}\sum_{<NN>}S_i\sigma_j - J^{\sigma-\sigma}_{<NN>}\sum_{<NN>}\sigma_j\sigma_k - J^{s-s}_{<NN>}\sum_{<NN>}s_i s_o$$
$$-J^{\sigma-\sigma}_{<NNN>}J\sum_{<NNN>}\sigma_j\sigma_l - J^{s-s}_{<NNN>}\sum_{<NNN>}S_i S_q - D\sum(S_i)^2, \quad [1]$$

donde los términos de $J^{a-b}_{<NN>}$ representan la interacción de intercambio a primeros vecinos de los espines "*a*" respecto de los espines "*b*", los términos $J^{c-c}_{<NNN>}$ representan la interacción de intercambio a segundos vecinos de los espines del mismo tipo. *D* representa al campo cristalino. Hemos utilizado para la simulación los siguientes valores:

$$J^{S-\sigma}_{<NN>} = -0.5;\ J^{S-\sigma}_{<NN>} = 0.0;\ J^{S-S}_{<NN>} = 0.2;\ J^{S-S}_{<NNN>} = 0.1;$$
$$J^{\sigma-\sigma}_{<NN>} = 1.0;\ J^{\sigma-\sigma}_{<NNN>} = 0.5;\ D = +1;\ S = 0,\pm 1\ y\ \sigma = \pm 1/2\ [2]$$

El espesor de cada una de las capas se mantiene fijo, se aumenta la concentración de espines $\sigma_x$ en la mezcla o capa mixta, con una distribución aleatoria y uniforme, en el rango $x = 0.000$ a $x = 0.60$. Utilizando el método de Monte Carlo con un barrido secuencial de la red, calculamos las magnetizaciones por sitio en la red de los espines *S*, $M_S$, y de los espines $\sigma$, $M_\sigma$, la magnetización total por sitio, *M* y la susceptibilidad magnética, $\chi$. De igual forma calculamos, la temperatura de compensación, $T_{comp}$, definida como la temperatura a la cual las magnetizaciones de las subredes tienen igual magnitud $\left|M_S(T_{comp})\right| = \left|M_\sigma(T_{comp})\right|$ y signos opuestos, de tal forma que la magnetización total es cero a $T_{comp}$.

## Resultados

Los resultados obtenidos para el modelo tridimensional $3(S_{1-x}\sigma_x)/\sigma$, representan el promedio de 30 configuraciones distintas, para cada valor de concentración de espines $\sigma_x$ Como se observa en la Figura 1, el aumento de $\sigma_x$ en la mezcla conduce a una disminución de la magnetización total del sistema y un desplazamiento de la temperatura de compensación. Cuando la concentración de espines $\sigma$ sobrepasa el 50% en la mezcla, la magnetización total se hace negativa. En la gráfica observamos que la temperatura crítica es independiente de la concentración de espines $\sigma$ en la mezcla.

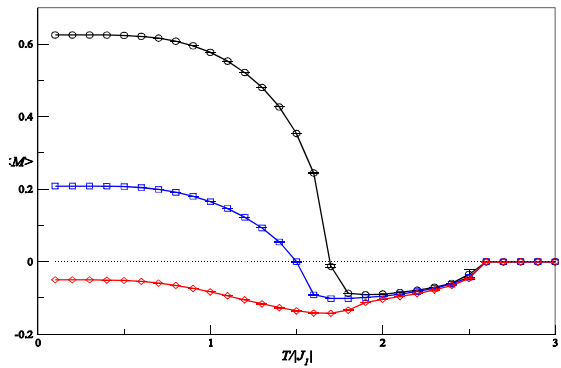

Figura 1. Magnetización total de las multicapas $3(S_{1-x}\ \sigma_x)/s$, para $x$= 0.00 (círculos), $x$= 0.30 (cuadrados) y $x$= 0.60 (diamantes).

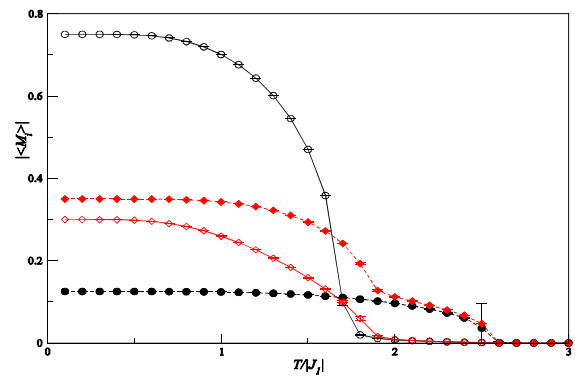

Figura 2. Módulo de las magnetizaciones de las subredes de las multicapas $S_{1-x}\ \sigma_x$, para $x$= 0.00 ($|<M_{S>}|$ círculos vacíos, $|<M\sigma>|$ círculos llenos) y $x$= 0.60 ($|<M_{s>}|$ diamantes vacíos,

En la Figura 2 graficamos los módulos de las magnetizaciones de las subredes para dos valores distintos de $x$, observando que para un aumento sustancial de $x$ ($x$= 0.60), la $T_{comp}$ desaparece. La introducción de forma aleatoria y uniforme de espines $\sigma_x$ en la capa mixta, es usada experimentalmente para disminuir la interdifusión en la interfaz de las capas. En general, la interdifusión en multicapas afecta las características del compuesto creado, por lo que su control es importante.

Haciendo un barrido en la concentración de espines $\sigma_x$, hemos determinado el comportamiento de la temperatura de compensación, como se muestra en la Figura 3. Se observa que la $T_{comp}$ disminuye lentamente con el incremento de $x$, hasta valores donde la cantidad de espines de un tipo y otro se hacen comparables (del orden del 40% de espines $\sigma$ respecto a un 60% de espi-

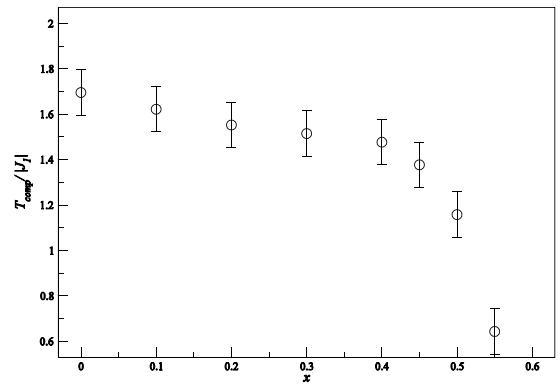

Figura 3. Comportamiento de la temperatura de compensación, $T_{comp}$, de la multicapa $3(S_{1-x}\ \sigma_x)/s$, con respecto al incremento de la nes $S$). Para una concentración superior al 40%, la $T_{comp}$ comienza a disminuir rápidamente hasta desaparecer. Los resultados encontrados en este trabajo coinciden cualitativamente con algunos resultados experimentales (6).

## Conclusiones

Hacemos un estudio numérico del comportamiento de la temperatura de compensación, , en multicapas ferrimagnéticas de tipo $3(S_{1-x} \sigma_x)/s$ con respecto al aumento de la concentración de espines $\sigma_x$, los cuales distribuidos de forma aleatoria y uniforme en la capa mixta $(S_{1-x} \sigma_x)$. Utilizamos el modelo mixto de Ising, ya que estamos considerando que tanto los espines $S = 0, \pm 1$ como los espines $\sigma = \pm 1/2$ se ordenan en estructuras son cúbicas, con interacciones de intercambio ferromagnéticas entre espines del mismo tipo y antiferromagnéticas entre espines distintos.

Encontramos que los parámetros de orden, magnetización total y magnetizaciones de los espines $S$ y $\sigma$ en el sistema, son susceptibles a cambios en la concentración de espines $\sigma_x$, observandose cambios en el comportamiento de las mismas.

El incremento de $x$ en el sistema cambia de un modelo ferrimagnético con punto de compensación a otro sin punto de compensación, pasando por un estado antiferromagnético. Este resultado es importante en problemas de interdifusión, donde la desaparición de la $T_{comp}$ en un sistema ferrimagnético, puede indicar una disminución en la interdifusión entre multicapas (7). Cuando el porcentaje de espines $s_x$ supera al porcentaje de espines $S$, la temperatura de compensación desaparece de forma abrupta, con lo que podemos establecer un valor crítico de $x$.

## Agradecimientos